\documentclass[oneside]{amsart}
\usepackage[T1]{fontenc}
\usepackage{geometry}
\makeatletter

\providecommand{\LyX}{L\kern-.1667em\lower.25em\hbox{Y}\kern-.125emX\@}
\newcommand{\noun}[1]{\textsc{#1}}

\theoremstyle{plain}    

\numberwithin{equation}{section} 
\numberwithin{figure}{section} 

\makeatother

\begin{document}

\title{Symplectic leaves of W-algebras from the reduced Kac-Moody point of view}

\author{Z. Bajnok and D. Nógrádi}

\address{\centering \emph{Institute for Theoretical Physics, Roland Eötvös University,
H-1117 Budapest, Pázmány sétány 1/A, Hungary }}

\begin{abstract}
The symplectic leaves of W-algebras are the intersections of the symplectic
leaves of the Kac-Moody algebras and the hypersurface of the second class constraints,
which define the W-algebra. This viewpoint enables us to classify the symplectic
leaves and also to give a representative for each of them. The case of the \( W_{2} \)
(Virasoro) algebra is investigated in detail, where the positivity of the energy
functional is also analyzed. 
\end{abstract}
\maketitle

\section{Introduction}

W-algebras have attracted a great interest since their first appearance \cite{GeDi}
thanks to the fact that their quantized versions \cite{FaLu}, (the extensions
of the Virasoro algebra with higher spin currents), are relevant not only in
the classification of two dimensional conformal field theories but also in describing
various statistical physical models. (For a review on W-algebras and their application
see \cite{BoSh} and references therein). Later it was shown in \cite{BaFe}
that the Toda models (which carry the W-algebras as symmetry algebras) are Hamiltonian
reductions of the Wess-Zumino-Witten (WZW) models. Under the reduction procedure,
which can be implemented by second class constraints, the symmetry algebra of
the WZW model namely the Kac-Moody (KM) algebra reduces to the symmetry algebra
of the Toda models, to the W-algebra. 

The quantization of W-algebras started by a free field construction \cite{FaLu},
then a BRST method \cite{deB} was adopted to produce they quantum counterparts.
None of the approaches mentioned however, relied on the classical geometry of
W-algebras loosing useful information in this way. The aim of this talk is to
reveal the classical geometry of W-algebras, more precisely to analyze their
symplectic leaves, which can be used to quantize them via geometric quantization
\cite{Wood}, later. Since this problem has its own mathematical relevance there
were some efforts in this direction previously. 

In the simplest case, which corresponds to the Virasoro algebra the program
mentioned means the investigation of the coadjoint orbits of the Virasoro group\cite{Wi}.
The classification of these orbits are well-known \cite{LaPa,Ki,Wi,OvKh,BFP,KhSh,GoJo}.
Although only in \cite{BFP} give the authors explicit representative for each
class. Moreover, they also investigated the positivity of the energy functional
on the orbits, which is relevant for finding the highest weight representations
at the quantum level. In the next simplest case, namely in the case of \( W_{3} \)
the orbits were classified in \cite{KhSh}, while the case of \( W_{n} \) was
considered in \cite{OvKh}. 

In my talk I will show that the symplectic leaves of W-algebras are the intersections
of the symplectic leaves of KM algebras and the hypersurface determined by the
second class constraints. This approach not only provides a unified viewpoint
for the orbit-classification obtained previously but also gives explicit representative
for each orbit. (I note that a Lagrangian realization of this idea was applied
in one particular case, namely in the case of some special orbits of the Virasoro
algebra in \cite{GoJo}). 

The talk is organized as follows. In Section 2 I give the definition of the
WZW phase space with its Hamiltonian and review how it is connected to the symplectic
leaves of its symmetry algebra the KM algebra. By means of these I classify
the symplectic leaves and provide a representative for each of them. In Section
3 I show how the W-algebras arise as reduction of the system above and how this
fact can be used in the classifying their symplectic leaves. At each stage the
results are demonstrated on the example of \( SL(2,\mathbb R) \).

\section{The phase space of the WZW model and the symplectic leaves of the KM algebra }

The Hamiltonian formulation of the WZW theories is the following. Take a maximally
non-compact Lie group \( G \) (this is necessary in order to carry out the
reduction) and define the phase space to be
\[
M_{WZW}=\left\{ (g,J)\vert \quad g\in LG,J\in L\mathfrak g\right\} \, \, ,\]
 where \( LG \) denotes the loop group of \( G \), that is smooth maps from
\( S^{1}=\{e^{ix}\vert \, \, x\in [0,2\pi )\} \) to \( G \) and \( L\mathfrak g \)
is its Lie algebra. It is equipped with the Poisson brackets:

\begin{eqnarray*}
\left\{ g(x),g(y)\right\} =0\, \, ,\quad \left\{ J_{a}(x),g(y)\right\} =t_{a}g(y)\delta (x-y) &  & \\
\left\{ J_{a}(x),J_{b}(y)\right\} =f_{ab}^{c}J_{c}(y)\delta (x-y)+k\kappa _{ab}\delta ^{\prime }(x-y) & ,
\end{eqnarray*}
where the current is decomposed as \( J(x)=J^{a}(x)t_{a}\: ;\, t_{a}\in \mathfrak g\, ,\, \, [t_{a},t_{b}]=f_{ab}^{c}t_{c} \),
and the indices are lowered and raised by the Cartan metric \( \kappa _{ab}=Tr(t_{a}t_{b}) \)
and its inverse \( \kappa ^{ab} \), respectively, and \( \delta (x-y) \) represents
the \( 2\pi  \) periodic delta function: \( \delta (x)=\frac{1}{2\pi }\sum _{n\in Z}e^{inx} \).
The second line contains the defining relations of the KM algebra at level \( k \).
If we introduce 

\[
\tilde{J}=-g^{-1}Jg+kg^{-1}g^{\prime }\, \, ,\]
 then the Hamiltonian takes the following simple form:
\[
H_{wzw}=\frac{1}{2k}\int _{0}^{2\pi }dxTr(J(x)^{2}+\tilde{J}(x)^{2})\, \, .\]
 In light cone coordinates, \( x_{\pm }=x\pm t \) , it generates the flow:
\begin{equation}
\label{emo}
\partial _{-}J(x,t)=0\quad ;\qquad k\partial _{+}g(x,t)=J(x,t)g(x,t)\, \, .
\end{equation}
 The most general solutions are \( J(x,t)=J(x_{+}) \) and \( g(x,t)=g(x_{+})g(x_{-}) \).
The symmetries of the model consist of local transformations, \( h(x_{+})\in LG \)
:

\[
g(x_{+})\to h(x_{+})g(x_{+})\quad ;\qquad J(x_{+})\to h(x_{+})J(x_{+})h^{-1}(x_{+})+kh^{\prime }(x_{+})h^{-1}(x_{+})\]
 and global transformations, \( h\in G \) :
\[
g(x_{+})\to g(x_{+})h\quad ;\qquad J(x_{+})\to J(x_{+})\, \, .\]
Observe that the local transformations are generated by the conserved KM currents
\( J(x_{+}) \) and are nothing but the coadjoint action of the centrally extended
loop group on its Lie algebra and left translation on itself. (Note that if
\( k\neq 0 \) what we will suppose in the sequel we can absorb \( k \) into
the definition of \( J \) , that is we change \( J \) to \( J/k \) and analyze
all the orbits in one turn by substituting \( k=1 \) in the formulae from now
on).

Now the key point in analyzing the coadjoint orbits is the fact, that there
is one-to-one correspondence between the currents \( J(x_{+}) \) and the elements
\( g(x_{+}) \) with the \( g(0)=e \) property via the equations of motions
(\ref{emo}). This means that instead of analyzing the coadjoint action on \( J(x_{+}) \)
we can analyze the much simpler action on \( g(x_{+}) \). Of course we have
to modify the action in order to ensure \( g(0)=e \). This can be achieved
by combining the local and global transformations in the following way: 
\[
g(x_{+})\to h(x_{+})g(x_{+})h(0)^{-1}\quad ;\qquad J(x_{+})\to h(x_{+})J(x_{+})h^{-1}(x_{+})+kh^{\prime }(x_{+})h^{-1}(x_{+})\, \, .\]
 According to this action we can split any transformation as 
\[
h(x_{+})=(h(0),h(x_{+})h^{-1}(0))\, \, ,\]
 where \( h(0) \) is a constant loop, \( h(x_{+})h^{-1}(0)\in \Omega G \)
and \( \Omega G \) consists of such loops which starts at the identity. 

Thus topologically we can write the loop group as \( LG=G\times \Omega G \).
The group structure is a semi-direct product: \( (a,g)(b,h)=(ab,g{\rm Ad}_{a}h) \). 

Remember that we have split the original periodic \( g(x,t) \) as \( g(x,t)=g(x_{+})g(x_{-}) \).
Consequently \( g(x_{+}) \) is not necessary periodic, only quasi periodic
\[
g(x_{+}+2\pi )=g(x_{+})M\quad ;\qquad M=g(2\pi )\, \, ,\]
 M is called the monodromy matrix. Under a loop group transformation \( g(x_{+})\to h(x_{+})g(x_{+})h(0)^{-1} \)
the monodromy matrix changes as \( M\to h(0)Mh(0)^{-1} \)which means that the
conjugacy class of the monodromy matrix is an invariant of the orbit. It is
in fact the only invariant, since if \( M_{1} \) and \( M_{2} \) are conjugated
\( M_{2}=hM_{1}h^{-1} \) and \( g_{i} \) corresponds to \( M_{i} \) then
\( g_{1}h^{-1}g_{2}^{-1} \) is periodic and maps \( g_{2} \) to \( g_{1} \).
Note however, that an orbit corresponding to a conjugacy class is not necessarily
connected. Its connected components are labelled by \( \Pi _{0}(LG)=\Pi _{1}(G) \)
that is by the fundamental group of \( G \). 

Concluding the connected components of the coadjoint orbits of the centrally
extended loop groups are characterized by the conjugacy class of the monodromy
matrix and the discrete invariant labelled by the fundamental group of \( G \). 

In order to analyze the topology of the orbit we need to determine the stabilizer
of a given \( J \). Clearly \( h(x_{+}) \) will stabilize a given \( J(x_{+}) \)
if and only if \( h(x_{+})g(x_{+})h(0)^{-1}=g(x_{+}) \) holds. This means that
\( h(0) \) must stabilize the monodromy matrix, \( h(0)\in H_{M}=\{g\in G\vert gMg^{-1}=M\} \),
moreover \( h\in G \) completely determines the stabilizing \( h(x_{+}) \)
as \( h(x_{+})=g(x_{+})hg(x_{+})^{-1} \). The topology of the orbit is 
\[
LG/H_{M}=G\times \Omega G/H_{M}=(G/H_{M})\times \Omega G\, \, ,\]
 where in the last equality we used the fact that \( LG/H_{M} \) is a principal
\( \Omega G \) bundle over \( H/M \) which turns out to be trivial.

\subsection{The example of the \protect\( SL(2,\mathbb R)\protect \) WZW model }

The coadjoint orbits are labelled by the conjugacy classes of \( SL(2,\mathbb R) \)
and by a \( \mathbb Z=\Pi _{1}(SL(2,\mathbb R)) \) valued discrete parameter.
For each conjugacy class we present a representative then give its stabilizer
group. Furthermore we determine that particular \( g(x_{+}) \) which gives
rise to this monodromy matrix and compute the corresponding \( J(x_{+})=g(x_{+})^{\prime }g^{-1}(x_{+}) \).
We proceed from case to case: 

\begin{enumerate}
\item Elliptic case:
\[
M=\left( \begin{array}{cc}
\cos (2\pi \omega ) & -\sin (2\pi \omega )\\
\sin (2\pi \omega ) & \cos (2\pi \omega )
\end{array}\right) \quad ;\qquad H_{M}=\left\{ \left( \begin{array}{cc}
\cos (\alpha ) & -\sin (\alpha )\\
\sin (\alpha ) & \cos (\alpha )
\end{array}\right) \quad \vert \, \, \alpha \in (0,2\pi ]\right\} \, \, ,\]
 
\begin{equation}
\label{ellip}
g(x_{+})=\left( \begin{array}{cc}
\cos (\omega x_{+}) & -\sin (\omega x_{+})\\
\sin (\omega x_{+}) & \cos (\omega x_{+})
\end{array}\right) \quad ;\qquad J(x_{+})=\left( \begin{array}{cc}
0 & -\omega \\
\omega  & 0
\end{array}\right) \, \, .
\end{equation}
These correspond to the leaf labelled by \( n=0 \). In order to move to the
\( n^{th} \) leaf we multiply \( g(x_{+}) \) by 
\[
T_{n}=\left( \begin{array}{cc}
\cos (nx_{+}) & -\sin (nx_{+})\\
\sin (nx_{+}) & \cos (nx_{+})
\end{array}\right) \, \, ,\]
which amounts to shift \( \omega  \) to \( \omega +n \) in (\ref{ellip}). 
\item Exceptional case 
\[
M=\eta \left( \begin{array}{cc}
1 & 0\\
0 & 1
\end{array}\right) \quad \eta =\pm 1\quad ;\qquad H_{M}=SL(2,\mathbb R)\, \, .\]
This can be recovered from the elliptic case by taking \( \omega =n\in \mathbb Z \)
for \( \eta =1 \) and \( \omega =n+1/2\in \mathbb Z+1/2 \) for \( \eta =-1 \). 
\item Hyperbolic case
\[
M=\eta \left( \begin{array}{cc}
e^{2b\pi } & 0\\
0 & e^{-2b\pi }
\end{array}\right) \quad \eta =\pm 1\quad ;\qquad H_{M}=\left\{ \left( \begin{array}{cc}
a & 0\\
0 & a^{-1}
\end{array}\right) \quad \vert \, \, a\neq 0\right\} \, \, .\]
 The corresponding \( g(x_{+}) \) and currents are
\[
g(x_{+})=T_{n}\left( \begin{array}{cc}
e^{bx_{+}} & 0\\
0 & e^{-bx_{+}}
\end{array}\right) \quad ;\qquad J(x_{+})=n\left( \begin{array}{cc}
0 & -1\\
1 & 0
\end{array}\right) +b\left( \begin{array}{cc}
\cos (2nx_{+}) & \sin (2nx_{+})\\
\sin (2nx_{+}) & -\cos (2nx_{+})
\end{array}\right) \, \, ,\]
where similarly to the previous case \( n\in \mathbb Z \) for \( \eta =1 \)
and \( n\in \mathbb Z+1/2 \) for \( \eta =-1 \). 
\item Parabolic case 
\[
M=\eta \left( \begin{array}{cc}
1 & 0\\
q & 1
\end{array}\right) \quad \eta =\pm 1\: ,\; q=\pm 1\quad ;\qquad H_{M}=\left\{ \left( \begin{array}{cc}
1 & 0\\
a & 1
\end{array}\right) \quad \vert \, \, a\in \mathbb R\right\} \, \, ,\]
\[
g(x_{+})=T_{n}\left( \begin{array}{cc}
1 & 0\\
\frac{qx}{2\pi } & 1
\end{array}\right) \quad ;\qquad J(x_{+})=n\left( \begin{array}{cc}
0 & -1\\
1 & 0
\end{array}\right) +\frac{q}{2\pi }\left( \begin{array}{cc}
\frac{1}{2}\sin (2nx_{+}) & -\sin ^{2}(nx_{+})\\
\cos ^{2}(nx_{+}) & -\frac{1}{2}\sin (2nx_{+})
\end{array}\right) \, \, .\]
Here, as before \( n\in \mathbb Z \) for \( \eta =1 \) and \( n\in \mathbb Z+1/2 \)
for \( \eta =-1 \). 
\end{enumerate}

\section{The definition of W-algebras and their symplectic leaves}

In order to give a self-contained description of W-algebras we collect the main
results of \cite{BaFe}. For simplicity we restrict ourselves to the case of
the \( SL(n,\mathbb R) \) WZW model. The generalization for other groups is
straightforward. Impose the following constraints on the KM current:
\[
\Phi _{\alpha }=J_{\alpha }-\chi (J_{\alpha })=0\quad \textrm{where }\, \, \chi (J_{\alpha })=\left\{ \begin{array}{l}
1\textrm{ if }\alpha \in \Delta _{-}\textrm{ }\\
0\textrm{ if }\alpha \in \Phi _{-}\setminus \Delta _{-}
\end{array}\right. \qquad J_{const}=\left( \begin{array}{ccccc}
* & \dots  & \dots  & \dots  & *\\
1 & * & \dots  & \dots  & *\\
0 & 1 & * & \dots  & *\\
\dots  & \dots  & \dots  & \dots  & \dots \\
0 & \dots  & 0 & 1 & *
\end{array}\right) \]
 where \( \Phi _{-} \) (\( \Delta _{-} \)) denotes the set of negative (simple)
roots of \( sl(n,\mathbb R) \). These constraints are first class so they generate
gauge transformations, which are nothing but the KM transformations generated
by the currents associated to the positive roots. One possible gauge fixing
is the so-called Wronsky gauge: 
\[
J_{gf}=\left( \begin{array}{ccccc}
0 & W_{2} & W_{3} & \dots  & W_{n}\\
1 & 0 & \dots  & \dots  & 0\\
0 & 1 & 0 & \dots  & 0\\
\dots  & \dots  & \dots  & \dots  & \dots \\
0 & \dots  & 0 & 1 & 0
\end{array}\right) \, \, ,\]
where the \( W \)-s are gauge invariant polynomials of the unconstrained KM
currents. The resulting phase space carries a Poisson algebra structure, which
is inherited from the WZW phase space and can be computed either by computing
the Poisson brackets of the gauge invariant quantities or by using the Dirac
bracket. The resulting Poisson algebra, which closes in general only on polynomials
of the fields
\[
\{W_{i},W_{j}\}=P_{ij}(W)\]
 is called the W-algebra associated to the group \( SL(n,\mathbb R) \) and
is denoted by \( W_{n} \). It always contains the Virasoro algebra:

\[
\{W_{2}(x),W_{2}(y)\}=-\frac{1}{2}\delta ^{\prime \prime \prime }(x-y)-2W_{2}(y)\delta ^{\prime }(x-y)+W_{2}^{\prime }(x)\delta (x-y)\]
 and there exists such a combination of the remaining generators that are primary
fields of weights \( 3,4,\dots ,n \) with respect to this Virasoro algebra.
Having fixed the gauge the equations of motions (\ref{emo}) forces \( g \)
to be of the form 
\[
g=\left( \begin{array}{cccc}
\psi _{1}^{(n-1)} & \psi _{2}^{(n-1)} & \dots  & \psi _{n}^{(n-1)}\\
\dots  & \dots  & \dots  & \dots \\
\psi _{1}^{\prime } & \psi _{2}^{\prime } & \dots  & \psi _{n}^{\prime }\\
\psi _{1} & \psi _{2} & \dots  & \psi _{n}
\end{array}\right) \, \, ,\]
where any of the \( \psi _{i} \) -s satisfies the following \( n^{th} \) order
differential equation: 
\[
\psi _{i}^{(n)}-W_{2}\psi _{i}^{(n-2)}-W_{3}\psi _{i}^{(n-3)}-\dots -W_{n-1}\psi ^{\prime }_{i}-W_{n}\psi _{i}=0\, \, .\]
Moreover, any infinitesimal W-transformation generated by \( Q_{i}=\int _{0}^{2\pi }dx\epsilon _{i}(x)W_{i}(x) \)
can be implemented by appropriately chosen field-dependent KM transformation
\( J_{imp}(W) \):
\[
\delta _{i}J_{gf}=\left( \begin{array}{ccccc}
0 & \{Q_{i},W_{2}\} & \{Q_{i},W_{3}\} & \dots  & \{Q_{i},W_{n}\}\\
0 & 0 & \dots  & \dots  & 0\\
0 & 0 & 0 & \dots  & 0\\
\dots  & \dots  & \dots  & \dots  & \dots \\
0 & \dots  & 0 & 0 & 0
\end{array}\right) =[J_{imp},J_{gf}]+J_{imp\, \, .}^{'}\]
The left hand side defines the tangent vector in the W-symplectic leaf which,
as a consequence of the KM implementation, is also a tangent vector on the coadjoint
orbit of the loop group. This shows that the W-symplectic leaves can be obtained
by considering the intersection of the coadjoint orbits of the loop group with
the gauge fixed surface. 

This view point allows one to classify the W-symplectic leaves: We have the
conjugacy class of the monodromy matrix as an invariant and the \( \Pi _{1}(G) \)
valued discrete topological invariant. This classification is not the finest
one however, since we also have to count the number of the connected components
of the intersection surface. We may not have any intersection, or we may have
one connected component or more than one. This analysis turns out to be very
complicated and we cannot cope with the general case, that is why we proceed
from case to case. Note however, that instead of analyzing the intersection
problem at the level of \( J \) we can analyze it in the language of \( g \)
(since we know that they are uniquely connected). This amounts to classify the
homotopy classes of nondegenerate curves in \( \mathbb R^{n} \) or equivalently
in \( \mathbb RP^{n-1} \). This problem was investigated in \cite{OvKh}.

\subsection{The example of the Virasoro algebra}

Let us focus on the \( SL(2,\mathbb R) \) case. Parameterizing the current
as \( J=\left( \begin{array}{cc}
J_{0} & J_{+}\\
J_{-} & -J_{0}
\end{array}\right)  \) the constraint reads as \( J_{-}=1 \) and the gauge fixed form is

\[
J_{gf}=\left( \begin{array}{cc}
0 & L\\
1 & 0
\end{array}\right) \quad ;\qquad L=J_{+}+J_{0}^{2}-J_{0\, \, .}^{\prime }\]
The Poisson bracket is simply
\[
\{L(x),L(y)\}=-\frac{1}{2}\delta ^{\prime \prime \prime }(x-y)-2L(y)\delta ^{\prime }(x-y)+L^{\prime }(x)\delta (x-y)\, \, ,\]
which is just the defining relation of the Virasoro algebra. Thus the coadjoint
orbits of the Virasoro group can be obtained by analyzing the intersection of
the coadjoint orbits of the loop group (previous chapter) and the gauge fixing
hypersurface. As in the case of the KM algebras it is transparent to work at
the level of \( g \). The gauge fixing forces \( g \) to be of the form of
\[
g=\left( \begin{array}{cc}
\psi _{1}^{\prime } & \psi _{2}^{\prime }\\
\psi _{1} & \psi _{2}
\end{array}\right) \quad ;\qquad \textrm{where }\quad \psi _{i}^{\prime \prime }-L\psi _{i}=0\]
 that is all the \( \psi _{i} \) -s satisfies the Hill equation. From the previous
chapter it follows that instead of analyzing the coadjoint orbits of the Virasoro
group we can analyze the conformally nonequivalent solutions of the Hill equation
or equivalently the homotopy classes of nondegenerate curves in the plane. 

It can be shown in the general case, that if the KM orbit has intersection with
the gauge fixed surface then the intersection is necessarily connected. In other
words we do not have new invariant compared to the KM case, and the only thing
we have to check for each connected orbit whether the intersection exists or
not. 

We can parameterize the representative of the KM currents listed in the previous
chapter for any leaf having \( n=0 \) as \( J_{0}=\left( \begin{array}{cc}
j_{0} & j_{-}\\
j_{+} & -j_{0}
\end{array}\right)  \) with constant \( j \)-s. The representative, \( J_{n} \), on the \( n^{th} \)
leaf can be obtained by acting with \( T_{n} \) as 
\begin{eqnarray*}
J_{n}= & T_{n}J_{0}T_{n}^{-1}+T_{n}^{\prime }T_{n}^{-1} & \\
= & \left( \begin{array}{cc}
j_{0}\cos (2nx_{+})-j_{1}\sin (2nx_{+}) & -n-j_{2}+j_{0}\sin (2nx_{+})+j_{1}\cos (2nx_{+})\\
n+j_{2}+j_{0}\sin (2nx_{+})+j_{1}\cos (2nx_{+}) & -j_{0}\cos (2nx_{+})+j_{1}\sin (2nx_{+})
\end{array}\right)  & .
\end{eqnarray*}
where \( j_{\pm }=j_{1}\pm j_{2} \). Now if \( R^{-2}=n+j_{2}+j_{0}\sin (2nx_{+})+j_{1}\cos (2nx_{+})>0 \)
then 
\[
h=\left( \begin{array}{cc}
R^{-1} & R^{\prime }(1+\frac{1}{n}R^{-2})\\
0 & R
\end{array}\right) \]
is periodic and is in the trivial homotopy class as a loop. It maps \( J_{n} \)
into the gauge fixed form and gives :
\[
L=C+2n(n^{2}+C)R^{2}+3n(n^{2}+C+2nj_{3})R^{4}\quad ;\qquad \textrm{where}\quad C=-\det J=j_{0}^{2}+j_{1}^{2}-j_{2}^{2}\, \, .\]
Now let us proceed form case to case: 

\begin{enumerate}
\item Elliptic case: intersection exists for \( n\geq 0 \) and we have the following
representative of the orbit
\[
L=-(n+\omega )^{2}\quad ;\qquad \omega \in (0,1)\quad \omega \neq \frac{1}{2}\]
 The stabilizer subgroup is \( S^{1} \). 
\item Exceptional case: can be recovered from the elliptic case by putting \( \omega =n \)
or \( \omega =n+1/2 \) where \( n\in \mathbb Z \). In the first case however,
only \( n>0 \) is allowed and the stabilizer is the whole \( SL(2,\mathbb R) \). 
\item Hyperbolic case: by conjugating the current \( J=b\left( \begin{array}{cc}
1 & 0\\
0 & -1
\end{array}\right)  \) with the constant \( \left( \begin{array}{cc}
1 & 0\\
b & 1
\end{array}\right)  \) one can arrange that \( R^{-2}=(n-1)+(b^{2}+1)\cos ^{2}(nx_{+})+(b\cos nx_{+}+\sin nx_{+})^{2}>0 \)
holds. The intersection exists for \( n\geq 0 \) and we have
\[
L=b^{2}+2n(n^{2}+b^{2})R^{2}+3n^{2}(b^{2}+2bn-n^{2})R^{4}\, \, .\]
 The stabilizer subgroup is \( \mathbb R \). 
\item Parabolic case: We have intersection for \( q=-1 \) if \( n>0 \) and for \( q=1 \)
if \( n\geq 0 \). Since \( R^{-2}=n+\frac{q}{2\pi }\cos ^{2}(nx_{+})>0 \)
the representative of the orbit is 
\[
L=n^{3}(2R^{2}-3(n+\frac{q}{2\pi })R^{4})\]
 and the stabilizer subgroup is \( \mathbb R \). 
\end{enumerate}
The positivity of the energy functional, \( E=\int _{0}^{2\pi }L(x_{+})dx_{+} \),
is necessary to obtain highest weight representation for the Virasoro algebra
since the energy in a quantum theory is bounded from below. 

In the local analysis we demand that \( \delta E=0 \) and \( \delta \delta E>0 \)
in order to have a minimum for the energy. Since 
\[
\delta _{\epsilon }E=\int _{0}^{2\pi }\epsilon (x_{+})\{L(x_{+}),L(y_{+})\}dy_{+}=-\int _{0}^{2\pi }\epsilon (y_{+})L^{\prime }(y_{+})dy_{+}\, \, ,\]
 (where we have dropped the total derivatives) it is clear that only constant
\( L \) can give local minimum of the energy. The second variation 
\[
\delta _{\epsilon }\delta _{\epsilon }E=\delta _{\epsilon }(E+\delta _{\epsilon }E)=-\int _{0}^{2\pi }(L(\epsilon ^{\prime })^{2}+\frac{1}{4}(\epsilon ^{\prime \prime })^{2})\]
shows that \( L\geq -\frac{1}{4} \) is necessary. The global analysis is much
more involved. The result is the following. The lowest energy is on the exceptional
orbit for \( \omega =1/2 \) with stabilizer \( SL(2,\mathbb R) \). So it is
a good candidate for the classical vacuum. The energy has a minimum also on
the elliptic orbits for \( n=0 \) and \( \omega <1/2 \). On the hyperbolic
orbit for \( n=0 \) it has a minimum. Moreover, surprisingly on the orbit corresponding
to \( q=-1 \) and \( n=1 \) in the parabolic case the energy is bounded from
below, however this lower bound is never reached. This indicates that the representation
obtained by quantizing this orbit is not of the highest weight type similarly
to the case for quantizing the cone-like coadjoint orbit of the group \( SL(2,\mathbb R) \).

\section*{Acknowledgements}

Z. B. thanks Ivailo Mladenov for the kind hospitality in Varna.

\end{document}